\input harvmac.tex
\figno=0
\def\fig#1#2#3{
\par\begingroup\parindent=0pt\leftskip=1cm\rightskip=1cm\parindent=0pt
\baselineskip=11pt
\global\advance\figno by 1
\midinsert
\epsfxsize=#3
\centerline{\epsfbox{#2}}
\vskip 12pt
{\bf Fig. \the\figno:} #1\par
\endinsert\endgroup\par
}
\def\figlabel#1{\xdef#1{\the\figno}}
\def\encadremath#1{\vbox{\hrule\hbox{\vrule\kern8pt\vbox{\kern8pt
\hbox{$\displaystyle #1$}\kern8pt}
\kern8pt\vrule}\hrule}}

\font\cmss=cmss10 \font\cmsss=cmss10 at 7pt
\font\cmss=cmss10 \font\cmsss=cmss10 at 7pt
\def\IZ{\relax\ifmmode\mathchoice
{\hbox{\cmss Z\kern-.4em Z}}{\hbox{\cmss Z\kern-.4em Z}}
{\lower.9pt\hbox{\cmsss Z\kern-.4em Z}} {\lower1.2pt\hbox{\cmsss
Z\kern-.4em Z}}\else{\cmss Z\kern-.4em Z}\fi}

\overfullrule=0pt

%
\def\cqg#1#2#3{{\it Class. Quantum Grav.} {\bf #1} (#2) #3}
\def\np#1#2#3{{\it Nucl. Phys.} {\bf B#1} (#2) #3}
\def\pl#1#2#3{{\it Phys. Lett. }{\bf B#1} (#2) #3}
\def\prl#1#2#3{{\it Phys. Rev. Lett.}{\bf #1} (#2) #3}
\def\physrev#1#2#3{{\it Phys. Rev.} {\bf D#1} (#2) #3}

\def\ijmp#1#2#3{{\it Int. J. Mod. Phys.} {\bf #1} (#2) #3}

\font\zfont = cmss10 

\def\bigone{\hbox{1\kern -.23em {\rm l}}}
\def\ZZ{\hbox{\zfont Z\kern-.4emZ}}

\def\a{\alpha}
\def\b{\beta}
\def\g{\gamma}
\def\d{\delta}
\def\e{\epsilon}

\def\k{\kappa}

\def\m{\mu}
\def\n{\nu}

\def\S{\Sigma}

\def\o{\over}

\Title{CALT-68-2209, hep-th/9901126}
{\vbox{
\hbox{\centerline{Instanton Action for Type II Hypermultiplets}}
}}
\smallskip
\centerline{Katrin Becker\footnote{$^\diamondsuit$}
{\tt beckerk@theory.caltech.edu} and Melanie 
Becker\footnote{$^\star$}
{\tt mbecker@theory.caltech.edu} }
\smallskip
\centerline{\it California Institute of Technology 452-48, 
Pasadena, CA 91125}
\bigskip
\baselineskip 18pt
\noindent
We analyze the hypermultiplet moduli space 
describing the universal sector of type IIA theory
compactified on a Calabi-Yau threefold.
The classical moduli space is described in terms
of the coset $SU(2,1)/U(2)$.
The flux quantization condition of the antisymmetric tensor field
of M-theory implies discrete identifications for the 
scalar fields describing this four-dimensional quaternionic 
geometry. 
Non-perturbative corrections of the classical theory
are described in terms of a membrane-fivebrane instanton
action which we construct herein.

\Date{January, 1999}
\newsec{Introduction}
Compactifications of type IIA superstring theory on a 
Calabi-Yau threefold results in a four-dimensional 
theory with $N=2$ supersymmetry.
Supersymmetry requires that the moduli space is a product space 
${\cal M}={\cal M}_V {\times}{\cal M}_H$, where ${\cal M}_V$ 
corresponds to the moduli space of vector multiplets and 
${\cal M}_H$ corresponds to the moduli space for hypermultiplets.
The vector multiplet moduli space is described in terms of a special 
K\"ahler manifold that has been well understood for some time.
The hypermultiplet moduli space is described in terms 
of a quaternionic geometry 
\ref\bw{J.~Bagger and E.~Witten, ``Matter Couplings in
$N=2$ Supergravity'', \np {222} {1983} {1}.}
whose quantum corrections turn
out to be more difficult to understand, in part
because of the complicated structure of these manifolds
\foot{A nice mathematical description can be found e.g. 
in 
\ref\as{P.~S.~Aspinwall, ``Aspects of the Hypermultiplet
Moduli Space in String Duality'', hep-th/9802194.}.}.
Quantum corrections to the classical 
hypermultiplet geometry have been
studied in some cases in the limit where gravity decouples
\ref\sewi{N.~Seiberg and E.~Witten, ``Gauge Dynamics and 
Compactification to Three Dimensions'', hep-th/9607163.}
\ref\ss{N.~Seiberg and S.~Shenker, ``Hypermultiplet
Moduli Space and String Compactification to Three Dimensions'', 
\pl {388} {1996} {521}, hep-th/9608086.}.
In this case the geometry is formulated in terms of hyper-K\"ahler 
geometry which turns out to be easier to understand.

Some time ago it was shown
that non-perturbative corrections 
to the hypermultiplet geometry of type II theories 
can be obtained from membrane and fivebrane instantons
\ref\bbs{K.~Becker, M.~Becker and A.~Strominger, ``Fivebranes, Membranes
and Non-Perturbative String Theory'', hep-th/9507158.}.
The explicit evaluation of some of these corrections was performed in the 
hyper-K\"ahler limit in
\ref\gmv{B.~R.~Greene, D.~R.~Morrison and C.~Vafa, 
``A Geometric Realization of Confinement'', \np {481} {1996} {513}, 
hep-th/9608039.} 
and
\ref\of{H.~Ooguri and C.~Vafa, ``Summing up D Instantons'', 
\prl {77} {1996} {3298}.}.
In this paper we would like to consider
the quantum moduli space of hypermultiplets
without decoupling gravity i.e. the full quaternionic geometry. 
This question is of interest since in some cases 
quantum corrections to the classical 
geometry are essentially gravitational in nature so that 
the hyper-K\"ahler limit becomes trivial.

We shall be interested in the quantum moduli
space for the so-called universal sector
\ref\cfg{S.~Cecotti, S.~Ferrara and L.~Girardello, 
``Geometry of Type II Superstrings and the Moduli Space of
Superconformal Field Theories'', \ijmp {4} {1989} {2475}.}
which appears
in every Calabi-Yau compactification of type II theories.
In
\ref\fesa{S.~Ferrara and S.~Sabharwal, ``Dimensional 
Reduction of Type-II 
Superstrings'', \cqg {6} {1989} L77.}
it was shown that the classical moduli space for
the universal hypermultiplet is
the coset $SU(2,1)/U(2)$.
A first step to understand the quantum  moduli space for 
hypermultiplets was done in  
\ref\strom{A.~Strominger, ``Loop Corrections to the Universal 
Hypermultiplet'', \pl {421} {1998} {139}, hep-th/9706195.}
\ref\afmn{I.~Antoniadis, S.~Ferrara, R.~Minasian and K.~S.~Narain, 
``$R^4$ Couplings in M and Type II Theories on Calabi-Yau Spaces, 
\np {507} {1997} {571}, hep-th/9707013.}.
These papers evaluated perturbative corrections to the classical 
hypermultiplet moduli space and 
showed that the moduli space receives a 
one-loop correction proportional to the Euler number of the 
internal Calabi-Yau.
This correction originates from the $R^4$-term of the 
{\cal M}-theory action of
\ref\gg{M.~B.~Green and M.~Gutperle, ``Effects of D Instantons'', 
\np {498} {1997} {195}, hep-9701093.}
\ref\gv{M.~B.~Green and P.~Vanhove, ``D-Instantons, 
Strings and M Theory'', 
\pl {408} {1997} {122}, hep-th/9704145.}.
Furthermore in {\strom} there appeared a proposal
for the correction to the classical metric to all orders
in perturbation theory. The corrected metric is 
related to the classical metric by a field redefinition.

In this paper we shall be interested in the explicit form 
of membrane and fivebrane 
instanton corrections to the classical
geometry. These non-perturbative corrections are encoded in
a membrane-fivebrane instanton action whose explicit form we
would like to compute. Such an action was calculated in
{\bbs} for the special case where the R-R background vanishes.
However, in order to describe the universal sector we are
interested in keeping the R-R fields coming from 
the eleven-dimensional
three-form field strength, so that a generalization of 
the results appearing
in {\bbs} is required.

Some mathematical aspects of our discussion overlap with 
a recent paper by O.~Ganor
\ref\Ga{O.~J.~Ganor, ``U-Duality Twists and Possible Phase
Transitions in $2+1$  Supergravity'', hep-th/9812024.}.
In section two we discuss the classical moduli space for the 
universal sector in some detail. The coset space 
$SU(2,1)/U(2)$ describing 
hypermultiplets has eight isometries which
leave the classical action invariant. 
The explicit form of these symmetry transformations was found in
\ref\wipr{B.~de Wit and A.~Van Proeyen, ``Symmetries
of Dual Quaternionic Manifolds'', \pl {252} {1990} {221}.}
and 
\ref\wi{B.~de Wit, F.~Vanderseypen and A.~Van Proeyen, 
``Symmetry Structure of Special Geometries'', 
\np {400} {1993} {463}, hep-th/9210068.}.
We will be particularly 
interested in three Peccei-Quinn symmetries corresponding
to shifts in the NS-NS axion and the two R-R three-form potentials. 
Charge quantization implies discrete
identifications for these fields. These transformations form a discrete
subgroup $Z$ so that the moduli space is
${\cal M}=Z{\backslash} SU(2,1)/U(2)$.
This discrete subgroup of the Peccei-Quinn symmetries is preserved
in the quantum theory.
Associated with the above isometries are a number of
conserved Noether charges.
The instanton action is expressed in terms of an
invariant combination of three
Noether charges.
In section three we construct the membrane-fivebrane instanton 
action explicitly. 
\newsec{Classical Moduli Space and Discrete Identifications}
Let us start by considering the classical moduli space for the 
universal sector.
\subsec{Eleven Dimensions}
The bosonic part of the eleven-dimensional supergravity action 
is\foot{Our signature is $(-,+,+,\dots, +)$.}
 \ref\cjs{E.~J.~Cremmer, B.~Julia and J.~Scherk, 
``Supergravity Theory in 11 Dimensions'', \pl {5} {1978} {409}.}:
\eqn\ai{ {\cal S}_{11}={1\o 2\k^2_{11}} \int d^{11} x \sqrt{-{\hat g} }
\hat R -{ 1\o 4 \kappa_{11}^2}
\int \left[\hat F_4 \wedge * \hat F_4  
-{1\o 3 } \hat A_3 \wedge \hat F_4 \wedge \hat F_4 \right]   ,  }
where $\hat A_3$ is the three-form potential, $\hat F_4=d\hat A_3$ is the
four-form field strength, $\k_{11}$ is the eleven-dimensional 
gravitational coupling constant and the `hat' denotes eleven-dimensional
quantities.
The field strength obeys the Bianchi identity
\eqn\di{ d{\hat F}_4=0,}
and the field equation
\eqn\dii{ d*{\hat F}_4+{1\o 2}{{\hat F}_4}^2=d \left( *{\hat F}_4+{1\o 2}
{\hat A}_3 \wedge {\hat F}_4 \right)=0. }
Equations {\di} and {\dii} give rise to two classically conserved charges
(see e.g. 
\ref\Page{D.N.~Page, ``Classical Stability of Round and Squashed
Seven-Spheres in Eleven-Dimensional Supergravity'', 
\physrev {28} {1983} {2976}.} and
\ref\guven{R.~G\"uven, ``Black p-Brane Solutions of
$D=11$ Supergravity Theory'', \pl {276} {1992} {49}.}).
The electric charge or `Page charge' 
\foot{We will be following the conventions of
\ref\dkl{M.~J.~Duff, R.~Khuri and J.~X.~Lu, ``String Solitons'',
hep-th/9412184.}.}
\eqn\diii{ q_e={1\o {\sqrt2}{\k}_{11}} {\int}_{S^7} \left( *{\hat F}_4+{1\o 2}
{\hat A}_3 \wedge {\hat F}_4 \right),  }
follows from the equation of motion for ${\hat F}_4$ and is the charge
associated to the membrane. Here $S^7$ is a seven-sphere surrounding
the membrane.
The magnetic charge or topological charge 
\eqn\div{ q_m={1\o {\sqrt2}{\k}_{11}} {\int}_{S^4} {\hat F}_4,  }
is the charge associated to the fivebrane. Here $S^4$ is an
asymptotic four-sphere
surrounding the fivebrane.
The electric and magnetic charges obey the Dirac quantization condition
(see {\dkl} and references therein)
\eqn\dv{ q_e q_m=2 \pi \IZ. }

These charges can be expressed in terms of the membrane and 
fivebrane tensions, which are functions of 
$\kappa_{11}$\ref\dlm{M.~J.~Duff, 
J.~T.~Liu and R.~Minasian,
``Eleven Dimensional Origin of String/String Duality: A One Loop Test'', 
\np {452} {1995} {261}.} 
\ref\dealwis{S.~P.~de Alwis, ``A Note on brane Tension and M Theory', 
\pl {388} {1996} {291}, hep-th/9607011.}.
To see this we need the form of the worldvolume action for the 
membrane in eleven dimensions. It is given by 
\eqn\aii{ 
\eqalign{ {\cal S}_{m}=T_2 \int d^{3} \xi \Bigl[ & -{1 \o 2} 
\sqrt{-\gamma} \gamma^{ij} \partial_i X^M \partial_j X^N {\hat g}_{MN}
+{1 \o 2} \sqrt{-\gamma}\cr
&+{1 \o 3!} \epsilon^{ijk} \partial_i X^M \partial _j X^N \partial_k X^P
{\hat A}_{MNP} (X) \Bigr] ,  }}
where ${\xi}^i$ (with  
$i=1,2,3$) are the worldvolume
coordinates, ${\g}^{ij}$ is the worldvolume metric, $X^M({\xi}^i)$ with
$M=0,\dots , 10$
describes the bosonic part of the membrane configuration and $T_2$ is 
the membrane tension.

Demanding the
membrane to have a 
well defined quantum theory, 
it follows from {\aii} that $\hat A_3$ has period $2\pi/T_2$. 
Therefore, $\hat F_4$ is quantized according to:
\eqn\aiii{ {\int}_{S^4}  \hat F_4={2\pi \o T_2} \IZ.  }
The consistency of the $\hat A_3$ periods
with the supergravity action {\ai} gives a relation between 
$\k_{11}$ and $T_2$
\eqn\aiv{ {2{\pi}^2 \o {\k^2_{11}T^3_2}}=\IZ .  }
This relation has been derived in the appendix
of \ref\dealwis{S.~P.~de Alwis, ``A Note on brane Tension and M Theory', 
\pl {388} {1996} {291}, hep-th/9607011.}.

The dual seven-form $*{\hat F}_4$ couples to the 
worldvolume of the eleven-dimensional
fivebrane. Demanding the fivebrane action to describe a well defined 
quantum theory gives a quantization conditions for the electric charge
\eqn\avx{ {\int}_{S^7} \left( *{\hat F}_4+{1\o 2}
{\hat A}_3 \wedge {\hat F}_4 \right)={2\pi \o T_5}\IZ.   }
Inserting {\aiii} and {\avx} in the Dirac quantization condition {\dv} 
one gets: 
\eqn\aviv{2 \kappa_{11}^2 T_2 T_5= 2\pi \IZ. }
Using {\aviv} and {\aiv} one obtains a relation
between membrane and fivebrane tensions which was first
derived by Schwarz
\ref\sch{J.~H.~Schwarz, ``The Power of M Theory'', hep-th/9510086.}
\eqn\aivxi{ T_5={1\o {2\pi}} T_2^2,  }
using duality arguments.
The expressions {\aiv} and {\aivxi} imply that membrane and fivebrane 
tensions can both be expressed in terms of the eleven-dimensional
gravitational coupling constant ${\k}_{11}$.

Due to the charge quantization conditions {\aiii} and {\avx}
the moduli space of the four-dimensional theory will have a set 
of discrete identifications or periodicities in the dual scalar
fields as we will see later on. 
To finish this section we would like to remark that 
equation {\dii} can be interpreted as the Bianchi identity
of the eleven-dimensional fivebrane.
As was argued in {\dlm} this equation will, in general, receive
gravitational Chern-Simons corrections associated with the sigma-model
anomaly on the six-dimensional fivebrane worldvolume.
The corrected fivebrane Bianchi identity contains a term proportional 
to $X_8$, which is an eight-form polynomial quartic in the gravitational 
curvature. This will give a gravitational contribution
to the quantization condition of the electic charge
{\diii}. We will not take this effect into account in the following.
The existence of gravitational contributions to the 
flux quantization law for the antisymmetric tensor field 
of M-theory was first pointed out in
\ref\witten{E.~Witten, ``On Flux Quantization in M-Theory
and the Effective Action'', hep-th/9609122.}.

\subsec{Ten Dimensions}
To reduce the action {\ai} to ten dimensions we 
make the following ansatz for the metric (see
\ref\ed{E.~Witten, ``String theory Dynamics in Various Dimensions'',
\np{443} {1995} {85}, hep-th/9503124.}
and references therein)
\eqn\av{ d{\hat s}^2_{11}=e^{-2\phi / 3} g_{mn}dx^mdx^n
+e^{4\phi/ 3} (dx^{11}-A_mdx^m)^2,}
where $m,n=1,\dots 10$ are ten-dimensional indices and  
$\phi$ is the ten-dimensional dilaton.
In the following we will not take the dependence on the type IIA
gauge field $A_m$ into account as this field contributes to the
four-dimensional vectormultiplet moduli space.
We will be using conventions where the range of the eleventh
dimension is
\foot{Here and in the following we will set $g_s=1$ for simplicity.}
$x_{11} \rightarrow x_{11}+2\pi \sqrt{\a'}$ and 
$2\k^2_{10}={\k^2_{11}/ {\pi \sqrt{\a'}}}=(2\pi)^7{\a'}^{4}$.
In order to obtain a canonical Einstein term in ten dimensions 
it is convenient to use the Weyl
rescaling formula (valid in any dimension $d$). Under a rescaling
of the metric of the form
\eqn\cvi{
{\tilde g}_{ac} = \Omega^{2} g_{ac},}
the scalar curvature term transforms as
\eqn\cv{
{\Omega}^{2-d}{\sqrt {\tilde g}}{\tilde R}=
 {\sqrt g}\Bigl[ R
 -(d-2)(d-1)g^{ac}(\partial_a \log \Omega)\partial_c \log\Omega  \Bigr]. }
After reducing to $d=10$ and
Weyl rescaling the metric with a factor
${\Omega}^2=e^{\phi/2}$ we obtain: 
\eqn\bv{ \sqrt{-{\hat g}}  {\hat R} \rightarrow 
{\sqrt {-g}} \left[ R -{1\o 2} \left( \partial_m \phi \right)^2 \right].  }
>From the eleven-dimensional 
three-form we can define ten-dimensional gauge field
potentials
\eqn\zi{\eqalign{
B_{mn} & = {\hat A}_{m n 11},  \cr 
A_{mnp} & ={\hat A}_{mnp}, }}
and ten-dimensional 
field strengths $H_3=dB_2$ and $F_4=dA_3$. 
After compactification and  Weyl rescaling we get
\eqn\avii{ 
{\hat F}_4 \wedge * {\hat F}_4 \rightarrow e^{\phi/2} F_4 \wedge * F_4 
+e^{-\phi} H_3 \wedge *H_3.
 }
The topological term is invariant under this rescaling and
has the form
\eqn\dx{{\hat A}_3 \wedge {\hat F}_4 \wedge {\hat F}_4
{\rightarrow } B_2 \wedge F_4 \wedge F_4 .
 }
To summarize, the 
ten-dimensional supergravity action takes the form
\eqn\avi{ \eqalign{&
S_{10}={1 \o 2 \kappa_{10}^2} \int d^{10}x \sqrt{-g} \Bigl[
R -{1 \o 2} (\partial_m \phi)^2 \Bigr]-{1\o 4 \kappa_{10}^2} \int
\Bigl[e^{\phi/2} F_4 \wedge * F_4 
+e^{-\phi} H_3 \wedge * H_3  \Bigr]\cr 
& +{1\o 4 \kappa_{10}^2} \int B_2 \wedge F_4 \wedge F_4.
} }

In ten dimensions a string is dual to a fivebrane and the
membrane is dual to a fourbrane. 
The string and the fourbrane are obtained by double dimensional 
reduction of the eleven-dimensional membrane and fivebrane respectively
\ref\tow{P.~K.~Townsend, ``D-Branes from M-Branes'', hep-th/9512062.}.
The quantization condition for the fivebrane charge follows from 
{\aiii} 
\eqn\dxi{{\int}_{S^3}H_3={2\pi \o T_1} \IZ,
 }
where $T_1=1/{2\pi\a'}$ is the string tension
which is related to the eleven-dimensional
membrane tension as $2\pi \sqrt{\a'} T_2=T_1$. 
The charge associated with the dual string follows from {\avx} and is 
quantized as 
\eqn\dxii{{\int}_{S^7}\left( e^{-\phi}*H_3+{1\o 2}A_3\wedge F_4
\right)={2\pi \o T_5} \IZ,
 }
where $S^7$ is a seven-sphere surrounding the string. 
Using {\aivxi} we can express the fivebrane tension in terms
of ${\a'}$ as $T_5=(2\pi)^{-5}{\a'}^{-3}$.
The 
fourbrane charge is quantized as 
\eqn\dxiii{{\int}_{S^4}F_4={2\pi \o T_2} \IZ,
 }
where the membrane tension is $T_2={({2\pi})^{-2}{\a'}^{-3/2}}$.
Finally, the dual membrane charge follows from {\avx}
\eqn\dxv{{\int}_{S^6}\left( e^{{\phi /2}}*F_4+A_3\wedge H_3
\right)={2\pi \o T_4} \IZ.
 }
The fourbrane tension can be obtained from the eleven-dimensional 
fivebrane tension as $T_4=2\pi{\sqrt{\a'}}T_5=(2\pi)^{-4}{\a'}^{-5/2}$.

Non-perturbative corrections to the four-dimensional universal 
hypermultiplet appear when the ten-dimensional type IIA 
membrane wraps a supersymmetric three-cycle and when the
fivebrane wraps the six-dimensional Calabi-Yau manifold.
The fivebrane charge {\dxi} and the membrane charge {\dxv}
give rise to four-dimensional Noether charges that are associated to
certain isometries in the four-dimensional hypermultiplet moduli space.
Due to the above charge quantization conditions a 
discrete subgroup of the four-dimensional isometries will be preserved
once instanton effects are taken into account.
We will see this in more detail in the following section.

{\subsec{Four Dimensions}
The complete action resulting from compactification of the type IIA 
theory on a 
Calabi-Yau threefold was computed in
\ref\fs{S.~Ferrara and S.~Sabharwal, ``Quaternionic Manifolds for Type II
Superstring Vacua of Calabi-Yau Spaces'', \np {332} {1990} {317}.}
\ref\bcf{M.~Bodner, A.C.~Cadavid, and S.~Ferrara, \cqg {8} {1991} {789}.}
\ref\ccaf{A.C.~Cadavid, A.~Ceresole, R.~D'Auria, and S.~Ferrara,
``Eleven-Dimensional Supergravity Compactified on 
Calabi-Yau Threefolds'', hep-th/9506144.}.
 Here we will follow the $SU(3)$-invariant reduction of
\ref\witten{E.~Witten, ``Dimensional Reduction of Superstring Models'', 
\pl {155} {1985} {151}.}, which was used by
Ferrara and Sabharwal {\fesa} 
to construct the manifold describing the universal sector 
of any Calabi-Yau compactification of type II theories.
The number of hypermultiplets in a four-dimensional 
Calabi-Yau compactification of the type IIA theory is $h_{21}+1$, 
while the number of vector multiplets is $h_{11}$. 
In the general case, with arbitrary $h_{11}$ and $h_{21}$
the moduli space is given by a sigma model with target manifold
described by the product of a K\"ahler manifold of complex dimension 
$h_{11}$ and a dual quaternionic manifold of complex dimension
$2(h_{21}+1)$. 
Compactifying
the type IIA theory on a manifold with 
$h_{21}=0$ yields only one hypermultiplet which 
is the so called `universal' hypermultiplet
{\cfg}.
The four bosonic fields in this multiplet are 
the dilaton, the NS-NS axion and two additional scalars of R-R type.
To perform the dimensional reduction to four dimensions one should keep 
only the $SU(3)$ singlets in the internal indices.
We will make the following ansatz for the metric
\eqn\aviii{ 
ds^2_{10}=e^{-\phi/2} g_{IJ}dx^Idx^J+
e^{3\phi /2}g_{\m\n}dx^{\m}dx^{\n},}
where $\mu,\nu=1,\dots,4$ are four-dimensional indices, 
$I,J=5,\dots ,10$ are six-dimensional indices 
and $\phi=\phi(x_{\mu})$. 
We shall be using complex coordinates to
describe the six-dimensional internal space. We 
will make an ansatz $g_{i \bar \jmath}=g_{\bar \jmath i}=
{\d}_{i \bar \jmath}$, with $i, \bar \imath =1,2,3$ 
for the internal metric.

The 
expectation value of the three-form potential 
is parametrized in terms of a complex R-R scalar $C$ as 
\eqn\avix{ A_{ijk}={\sqrt 2}C\e_{ijk}.}
The remaining fields contributing to the
`universal' hypermultiplet are the NS-NS tensor field
$B_{\m\n}$ with field strength $H_3=dB_2$ and the dilaton $\phi$.
In order to obtain the scalar manifold describing 
the four-dimensional quaternionic geometry
one has to integrate $H_3$ out.
This can be done by adding to the four-dimensional
action resulting from compactification of {\avi} a
Lagrange multiplier
\eqn\axi{ \eqalign{
S_4= & {1 \o 2 \kappa_{4}^2} \int d^{4}x \sqrt{-g} \left[
R -2(\partial_{\m} \phi)^2
-2e^{2\phi}|\partial_{\m} C|^2
\right]\cr &  -{1\o 4 \kappa_{4}^2} \int \left[e^{-4\phi} H_3 \wedge * H_3
- 2i H_3 \wedge {\bar C} 
{\buildrel \leftrightarrow \over d } C- 4 H_3 \wedge  dD \right].\cr }}
Here $\k_4^2=\k_{10}^2/(2\pi \sqrt{\a'})^6={\pi}{\a}'$    
is the four-dimensional gravitational 
constant.
We can dualize $H_3$ in terms of the pseudoscalar $D$ and
the complex field $C$ as 
\eqn\axii{ H_3=e^{4\phi}*\left(2d D+i  {\bar C}
{\buildrel \leftrightarrow \over d}
C\right).}
After integrating $H_3$ out 
the dual action describing the scalar fields takes the form
\eqn\ax{
S_4=-{ 1 \o \kappa_4^2} \int d^4 x \sqrt{-g}
\left[ \left(\partial_{\mu} \phi \right)^2 + e^{2 \phi}     
\mid \partial_{\mu} C \mid^2 +e^{4 \phi}  \left(\partial_{\mu} D 
+ {i \o 2}
{\bar C}  {\buildrel \leftrightarrow \over \partial_{\mu} } C \right)^2
\right].
}

We can introduce a new complex field $S$
\eqn\axiii{ S=e^{-2{{\phi} }}+2iD
+C\bar C.}
In terms of the complex fields $C$ and $S$ the classical Lagrangian
{\ax} can be written as
\eqn\axiv{{\cal L}=-{1 \o \kappa_4^2}\left( {\cal K}_{,S{\bar S}}
{\partial}_{\m}S{\partial}^{\m}{\bar S}
+{\cal K}_{,S{\bar C}}{\partial}_{\m}S{\partial}^{\m}{\bar C}
 +{\cal K}_{,C{\bar S}}{\partial}_{\m}C{\partial}^{\m}{\bar S}
 +{\cal K}_{,C{\bar C}}{\partial}_{\m}C{\partial}^{\m}{\bar C}\right).}
Here ${\cal K}$ is the 
K\"ahler potential, which has the Fubini-Study form
\eqn\axv{ {\cal K}=-log(S+\bar S-2C\bar C).  }
Notice that $e^{2 \phi}=2e^{\cal K}$.
The line element is explicitly
\eqn\e{ ds_4^2=e^{2{\cal K}}\left(dSd{\bar S}-2CdSd{\bar C}
-2{\bar C}d{\bar S}dC+2(S+{\bar S})dCd{\bar C} \right).}
This is a quaternionic manifold of real dimension four
corresponding to the coset space $SU(2,1)/U(2)$ {\fesa}.
The explicit form of the eight symmetry transformations which leave the 
classical Lagrangian invariant was found in {\wipr} and {\wi}.
Four classical symmetries have the form
\eqn\axvvvxii{ 
\eqalign{
S & \rightarrow S-{\epsilon}_0S-{i \o 4} {\epsilon}_1S^2-{1 \o 2}
({\epsilon}_3+i{\epsilon}_4)CS,  \cr 
C &  \rightarrow C-{{\epsilon}_0 \o 2}C-{i \o 4}{\epsilon}_1CS-
{1 \o 2}{\epsilon}_3(C^2-{S \o 2})-{i \o 2}{\epsilon}_4(C^2+{S \o 2}). \cr}}
Here the epsilons are real parameters. In particular, ${\epsilon}_0$
parametrizes a scale transformation.
We expect that these symmetries are not preserved in the quantum theory.

The situation is rather different for the remaining four symmetries. 
First, a duality symmetry exchanges
$ReC$ and $ImC$. We will later see that this symmetry 
is present when instantons are taken into account. 
Furthermore, 
there are three isometries associated with 
constant shifts of the NS-NS axion $D$ and the 
two R-R scalars $C$ and ${\bar C}$
{\fs}
\eqn\axvvvx{ 
\eqalign{
S & \rightarrow S+  i \a + 2({\g}+i{\b})C+{\g}^2+{\b}^2,  \cr 
C &  \rightarrow C  +{\g}-i{\b}, \cr}}
under which the classical Lagrangian is invariant.
Here $\a$, $\b$ and $\g$ are real paramaters.
The generators of these transformations form a continuos
nonabelian group $G$ known as the Heisenberg group.
The generators of $G$ obey the commutation relations
\eqn\eix{ [T_{\a},T_{\b}]=[T_{\a},T_{\g}]=0 \quad and \quad
[T_{\b},T_{\g}]=T_{\a}.}
It is expected that a discrete subgroup of these three symmetry
transformations remains in the quantum theory.
The reason for this is as follows.
Associated with the three isometries {\axvvvx} 
are
three classically conserved Noether currents. These are
\eqn\dxx{ 
\eqalign{
J_{\a} & ={i \o {\k}_4^2}
e^{2{\cal K}}\left(
dS-d{\bar S}+ 2 C 
 {\buildrel \leftrightarrow \over d } {\bar C} 
\right), \cr 
J_{\b} & =
-{ 2 i \o \kappa_4^2} e^{\cal K}\left(d C -d{\bar C} \right)
+2(C+{\bar C}) J_{\a} , \cr 
J_{\g} & =
-{ 2  \o \kappa_4^2} e^{\cal K} \left( dC +d{\bar C} \right)
-2i(C-{\bar C}) J_{\a} 
.\cr}} 
There are three corresponding conserved charges
\eqn\dxxi{{Q}_{\a,\b,\g}={\int}_{{\S}_3}*{ J}_{\a,\b,\g}, }
where ${\S}_3$ is a three cycle.
The charge $Q_{\a}$ corresponds to the fivebrane
charge {\dxi} while $Q_{\b}$ and $Q_{\g}$ correspond to membrane charges
coming from {\dxv}. There are two membrane charges because there are 
two homology classes for three cycles.

A general membrane-fivebrane instanton 
in four dimensions is described by the Euclidean  
continuation of the three
charges $(Q_{\a},Q_{\b},Q_{\g})$.
The charge quantization conditions will imply that only a discrete
subgroup of the symmetry transformations ${\axvvvx}$ will remain 
once instanton effects are taken into account. 
The discrete identification for the pseudoscalar $D$ 
can be derived from the quantization condition 
for the electric charge which follows from {\dxii}
\eqn\axx{ D \rightarrow D+{n_{\a} \o 2}+\dots, }
where $n_{\a}$ is an integer.
The field $S$ has then the periodicity
\eqn\axxi{ S \rightarrow S+in_{\a}+\dots. }
The periodic identification for the R-R scalar $C$ follows from
the quantization condition of the magnetic charge {\dxiii}
\eqn\axviii{ C \rightarrow C+n_{\g}-in_{\b},  }
where $n_{\b}$ and $n_{\g}$ are both integers.

To summarize, periodicity in the fields following from
R-R charge quantization conditions implies that we must identify under the 
action of a discrete non-abelian subgroup $Z$ of $G$
\eqn\axxx{ 
\eqalign{
S & \rightarrow S+  i n_\a + 2(n_{\g}+in_{\b})C+
n_{\g}^2+n_{\b}^2,  \cr 
C &  \rightarrow C  +n_{\g}-in_{\b}. \cr}} }
The resulting moduli space is then ${\cal M}=Z{\backslash} SU(2,1)/U(2)$.

The charges {\dxxi} are not single valued on the moduli space.
Under the discrete identifications 
{\axviii} and {\axxi} they transform as
\eqn\dxxii{{Q}_{\a}\rightarrow {Q}_{\a} \quad
{Q}_{\b} \rightarrow{Q}_{\b}+4n_{\g}{Q}_{\a}
\quad and \quad {Q}_{\g}\rightarrow {Q}_{\g}-4n_{\b}{Q}_{\a}.}
This interesting non-abelian structure is due to the presence 
of the $\int {\hat A}_3 \wedge {\hat F}_4$ term in the
eleven-dimensional electric charge {\diii}.
This structure will enable us to evaluate instanton effects 
coming from eleven-dimensional membranes and fivebranes 
using a rather simple argument, as we shall later see.

The existence of multi-valued charges may sound strange at first. 
This type of behavior was first
discussed by Witten \ref\witt{E. Witten, 
``Dyons of Charge $e \theta/ 2 \pi$'', \pl {86} {1979} {283}.} for 
CP-violating field theories containg axions and monopoles.
According to Witten a magnetic monopole in a theta vacuum 
becomes a dyon with an electric charge proportional to theta
\foot{A nice discussion on 
nonabelian vortices can be found in
\ref\preskill{H.~K.~Lo and J.~Preskill, 
''Non-Abelian Vortices and Non-Abelian Statistics'', 
\physrev{48}{1993}{4821}.}.}. 
In the context of string theories 
an analog situation was discussed by
Greene, Shapere, Vafa and Yau in relation to the 
``stringy cosmic string''\ref\scs{B.~R.~Greene, A.~Shapere, C.~Vafa 
and S.~T.~Yau, ``Stringy Cosmic string and 
Noncompact Calabi-Yau Manifolds'', \np {337} {1990}{1}.}.
Here it was noticed that if one followed certain string states 
adiabatically around closed loops they would not come back 
as the same state.
This is precisely what happens in our context. An element 
of $Z$, characterized by three integers $(n_{\a}, n_{\b}, n_{\g})$
can be associated to every closed loop in spacetime.
These integers are not invariant under $Z$ transformations because
the generators of $Z$ obey the non-abelian Heisenberg algebra.
Since a string in four dimensions is surrounded by a closed loop,
an element of $Z$ is associated to every string.
A fundamental string has $(n_{\a}, n_{\b}, n_{\g})=(\pm 1, 0,0)$.
This particular element of $Z$ is invariant under all $Z$ transformations 
so the notion of a fundamental string is globally defined.
The other elements of $Z$ are carried by strings which may be 
described as fourbranes wrapping supersymmetric
three-cycles in one of the two homology classes.
When these strings are dragged around one another they pick up 
fundamental string charge.

Invariant charges can be defined as
\eqn\dxxiii{ {\hat {Q}}_{\a}={Q}_{\a} \quad 
 {\hat {Q}}_{\b}={Q}_{\b}-4{\zeta}_{0}{Q}_{\a}
\quad and \quad  {\hat {Q}}_{\g}={Q}_{\g}-
4{\tilde{\zeta}}_{0}{Q}_{\a}.}
Here we have defined $C=\zeta+i{\tilde \zeta}$ and the 
subindex indicates the value of the field at infinity.

The classical action for the universal sector  
could in principle receive both perturbative 
and non-perturbative corrections in the quantum theory.
Perturbative corrections 
have been discussed in
{\strom} and {\afmn}.
These papers showed that the $R^4$-term of the 
M-theory action of {\gg} and {\gv} gives a one-loop
correction to the hypermultiplet metric which is proportional 
to the Euler number of the internal Calabi-Yau
\eqn\dxxia{{\chi}=2(h_{11}-h_{21}).}
Furthermore in {\strom} there appeared a proposal 
for a perturbative all-orders corrected metric 
which is related to the classical metric by a field 
redefinition.
In the following we will discuss
non-perturbative corrections to the classical moduli space
of the universal hypermultiplet.

\newsec{Non-Perturbative Corrections}

\subsec{Four-Fermi Coupling and Quaternionic Geometry}

In the following we would like to compute instanton 
corrections to the low energy effective action described by {\axv}.
Recall that in the context of $N=2$, $D=4$ supergravity
it was shown in 
{\bw} that the 
$4n$ scalars of $n$ hypermultiplets describe a quaternionic
geometry with holonomy group $Sp(n)Sp(1)$ and a 
non-vanishing $Sp(1)$ connection.
The Riemann tensor of this geometry can be written 
in the form
\eqn\ddi{ R_{ijkl}{\g}^l_{CI}{\g}^k_{BJ}={\epsilon}_{CB}R_{ijIJ}+
{\epsilon}_{IJ}R_{ijCB}, }
where $C,B=1,2$ $I,J=1,\dots , 2n$ and $i,j=1, \dots , 4n$;
$R_{ijAB}$ and $R_{ijIJ}$ are the $Sp(1)$ and $Sp(n)$ 
curvatures respectively and ${\g}^i_{AJ}$ are covariantly
constant functions of the $4n$ scalars that satisfy identities
similar to those of Dirac gamma matrices.
The $Sp(1)$ connection can be written in the form
\eqn\ddii{ R_{ijAB}={\k}^2 \left( {\g}_{iAI}{\g}_{jB}^I
-{\g}_{jAI} {\g}_{iB}^I\right), }
and the $Sp(n)$ connection is given by
\eqn\ddiii{ R_{ijIJ}={\k}^2 \left( {\g}_{iAI}{\g}^A_{jJ}
-{\g}_{jAI} {\g}^A_{iJ}\right)+{\g}^{AL}_i{\g}^K_{jA}R_{IJKL}, }
where $R_{IJKL}$ is totally symmetric in its indices and 
$\k^2$ is proportional to the four-dimensional gravitational constant.
In the following we will be computing non-perturbative
corrections to the four-fermi coupling {\bw}
\eqn\ddiv{\int d^4x {\sqrt g} ({\bar \chi}^I{\chi}^J)({\bar \chi}^K{\chi}^L)
R_{IJKL}
, }
where ${\chi}^I$ is the fermionic component of the universal hypermultiplet.
As noticed in {\bbs}
this correction implies a non-perturbative correction to the classical metric
on the moduli space
by $N=2$ supersymmetry.

\subsec {Membrane-Fivebrane Instanton Action}
There are two types of 
non-perturbative (instanton) corrections to the
$SU(2,1)/U(2)$ geometry described by {\axv}.
These corrections originate from the ten-dimensional Euclidean
fivebrane wrapping the Calabi-Yau space and from the ten-dimensional
Euclidean membrane wrapping a non-trivial Calabi-Yau three-cycle.
The type IIA fivebrane can be 
described in terms of an exact
conformal field theory 
\ref\cft{C.~Callan, J.~A.~Harvey and A.~Strominger,
``Worldbrane Actions for String Solitons'', 
\np {367} {1991} {60}.}.
The instanton action for the fivebrane was computed in {\bbs} 
for the special case
when
 the R-R background $C$ vanishes. To describe a correction 
to the universal hypermultiplet we are interested in keeping the 
R-R scalar $C$, so that a generalization of the
result {\bbs} is in order.
Let us recapiltulate the main points of this calculation.
Closely related are computations appearing  
in 
\ref\gist{S.~B.~Giddings and A.~Strominger, ``
String Wormholes'', \pl {230}{1989} {46}.} 
and the D-instanton calculations of
\ref\gg{M.~B.~Green and M.~Gutperle, ``Effects
of D-Instantons'', hep-th/9701093.}.
The basic idea used in {\bbs} to compute the fivebrane
instanton action is that the type IIA fivebrane 
can be described as a soliton {\cft}. Since the 
internal six-dimensional geometry is unaffected by the fivebrane 
instanton, we used standard four-dimensional instanton
methods to compute the instanton action. 

The fivebrane soliton is a solution to the ten-dimensional
field equations which is asymptotically flat in the four transverse 
directions.
We will make the ansatz {\aviii} for the ten-dimensional metric.
The relevant terms  
of the Euclidean four-dimensional action are:
\eqn\avid{ S_{4}=
{1 \o 2 \kappa_{4}^2} \int d^{4}x \sqrt{g} \Bigl[
-R +2 (\partial_{\m} \phi)^2 \Bigr]+{1\o 4 \kappa_{4}^2} \int
e^{-4\phi} H_3 \wedge * H_3,
}
or in terms of the dual field $D$ {\axii}:
\eqn\wwi{
S_4={ 1\o 2\kappa_4^2} \int d^4 x \sqrt{g} 
\Bigl[-R+2 (\partial_{\mu} \phi)^2 +2  
e^{4\phi} 
(\partial_{\mu} D)^2 \Bigr]
.}
If one considers variations of {\avid} which vanish on the boundary 
one obtains the Euclidean equations of motion
\eqn\ddvix{
\eqalign{
& R-2 (\partial_{\mu} \phi)^2 +{1 \o 12} e^{-4 \phi} H_3^2=0, \cr
& {\partial}_{\m}^2\phi+{1\o 12} e^{-4 \phi} H_3^2=0. \cr}}
The `neutral' solution describing the fivebrane soliton is
\eqn\ddv{
\eqalign{
& g_{\m\n}=\delta_{\m\n} e^{-2\phi}, \cr 
& e^{2 \phi}= e^{2{\phi}_0}+{ \alpha' n_{\a} \o y^2}, \cr
& H_{\m\n\rho}=2 {\epsilon_{\m\n\rho}}^{\lambda}
 \partial_{\lambda} \phi.\cr 
}}
Here $y^2={\d}_{\m\n}y^{\m}y^{\n}$ is the distance in the 
four-dimensional transverse space. 
If we consider variations of $D$ that do not vanish on the 
boundary, we have to add to the action the boundary term: 
\eqn\wwii{
-2i {\oint}_{\partial M} d {\S}^{\mu}  J_{\mu}^{\alpha} D,} 
in order to make it stationary. Here $J_{\mu}^{\alpha}$ represents the
Euclidean continuation of the current appearing in {\dxx} with $C=0$. 
This boundary term plays a similar role as the $\theta$-term 
appearing in conventional Yang-Mills theories. 
We have restored the $g_s$ dependence of the 
instanton action by comparing with the $C=0$ result of {\bbs}
where a careful analysis of the $g_s$ dependence was performed.

Taking into account the 
field configuration {\ddv} as well as the form of the four-dimensional
action {\avid} plus the boundary term, 
we obtain the instanton action
\eqn\ddvxi{S_{inst}=-Q_{\alpha} \left( {{1\o g_s^2}}+2i D_0
\right)=-2 \pi n_{\a} S_0,  }
where $g_s=e^{\phi_0}$ and $D_0$ are the asymptotic values of the dilaton 
and the axion respectively. Consequently, the field $S_0$ 
is the value of the field {\axiii} at infinity
(again for $C=0$). 
 
This instanton couples to the charge $Q_{\a}$ as this charge 
is related to $H_3$ and the fivebrane is a source for $H_3$.
>From the counting of fermionic zero modes it was
shown in {\bbs} that the above instanton action gives a
non-perturbative correction to the coupling of four dilatinos.
By supersymmetry such a correction is related to a
non-perturbative correction of the $S-{\bar S}$ component of the
classical metric on the moduli space.
The $e^{-1/g_s^2}$ dependence is typical for ordinary
Yang-Mills instantons. 

As we had already mentioned, a general membrane-fivebrane instanton is 
characterized by the three 
charges {\dxxi}. The charges 
$Q_{\b}$ and $Q_{\g}$ correspond to membrane charges.
The exponential of the 
instanton action should be invariant under
the discrete identifications {\axviii} and {\axxi}.
This constraint together with the result {\ddvxi} for $C=0$ 
uniquely fixes the action for an instanton with 
charges $(Q_{\a}, Q_{\b}, Q_{\g})$ as
\eqn\ddvi{{\cal S}_{inst}=-{1\o g_s^2}{\hat Q}_{\a}
-{1 \o g_s}
\left( {\hat Q}_{\b}+{\hat Q}_{\g}\right)-
i\left(2D_0Q_{\a}+{{\tilde{\zeta}}_0\o 2}Q_{\b}+{{\zeta}_0 \o 2}Q_{\g}\right).}
As is by now well known, membrane instantons, as opposed to 
fivebrane instantons, give
$e^{-1/g_s}$ corrections to the classical action.
This is the origin for the different $g_s$ dependence
for the bulk terms of the action which contain the membrane
charges.
Notice that this action is invariant under the duality 
symmetry which exchanges $\zeta$ and $\tilde \zeta$.
As argued in {\bbs} the counting of Goldstino zero modes 
in the instanton background tells us that the above action
contributes to a correction to the four-fermi interaction
which by supersymmetry is related to a correction of the 
metric on the moduli space appearing in {\e}. 
However, we have not worked out the explicit
form of the metric on the moduli space once these
instanton effects have been taken into account.
We hope to report on more details elsewhere.

Taking this computation as a guiding principle 
other more complicated quaternionic geometries could be understood 
along the same lines.
A more complicated example of quaternionic geometry 
was discussed by Bodner and Cadavid
\ref\boca{M.~Bodner and A.~C.~Cadavid, ``Dimensional Reduction
of Type IIB Supergravity and Exceptional Quaternionic
Manifolds'', \cqg {7} {1990} {829}.} in the context of 
type IIB compactifications.
Compactifying the type IIB theory on a Calabi-Yau manifold
with $h_{11}=1$ and $h_{21}=0$ yields a four dimensional theory
described in terms of two hypermultiplets. The classical moduli
space of these hypermultiplets is a quaternionic
manifold of real dimension eight which is the coset space
$G_{2,2}/SO(4)$. Equivalently, by mirror symmetry this
manifold can be obtained by compactifying the type IIA
theory on a Calabi-Yau manifold with $h_{21}=1$.
The instanton action describing non-perturbative corrections
of this geometry can be computed with similar methods as
we have done in this paper.
Work in this direction is in progress
\ref\kbmb{K.~Becker and M.~Becker, Work in progress.}.

\newsec{Conclusion}
The universal hypermultiplet is the first 
quaternionic manifold appearing in a four-dimensional
Calabi-Yau compactification whose perturbative and 
non-perturbative corrections have been understood
in some detail.
These corrections cannot be evaluated in the simpler 
hyper-K\"ahler limit, as is usually done in the literature. 
This simple but important example may teach us how to 
explicitly compute the quantum moduli space of a general quaternionic
geometry.
Our undertanding of vector multiplet and hypermultiplet
moduli spaces in the context of
type II compactifications would then finally be on equal footing.

\vskip 0.5cm

\noindent {\bf Acknowledgement}

\noindent
We thank Andy Strominger for his collaboration on some of these issues
some time ago and John Schwarz for discussions. 
This work was supported by the U.S. Department of Energy 
under grant DE-FG03-92-ER40701.   

\vskip 0.5cm
\noindent {\bf Note Added}

\noindent Some mathematical aspects of
this paper overlap with a recent paper by O.~Ganor
{\Ga}.

\listrefs

\end